\documentclass[aps,prb,twocolumn,amsmath,amssymb,superscriptaddress,reprint,floatfix]{revtex4-2}
\usepackage{upgreek}
\usepackage{graphicx}
\usepackage{commath}
\usepackage{bm}
\usepackage{dsfont}
\usepackage[usenames,dvipsnames]{xcolor}
\usepackage{pstricks}
\usepackage[tight]{subfigure}
\usepackage{verbatim}
\usepackage{units}
\usepackage{multirow}
\usepackage{mathrsfs}
\usepackage{leftidx}
\usepackage{xspace}
\usepackage{braket}
\usepackage{bbm}
\usepackage{amsfonts,amssymb,amsmath}
\usepackage{psfrag}
\usepackage[abs]{overpic}
\usepackage{hyperref}
\hypersetup{colorlinks=true,linktoc=all,linkcolor=blue,breaklinks=true,citecolor=blue,urlcolor=blue}
\usepackage{color}
\usepackage{caption}
\usepackage{float}
\usepackage{booktabs}
\usepackage{multirow}

\usepackage{lineno}  
\usepackage{braket}
\usepackage{booktabs}
\usepackage{mathtools}
\usepackage{makecell}
\usepackage[version=4]{mhchem}
\usepackage{times}
\usepackage{mathptmx}
\usepackage{footnote}
\usepackage{amsmath}

\begin{document}
\captionsetup[figure]{labelfont={bf},labelsep=period,name={Figure}}
\captionsetup[table]{labelfont={bf},labelsep=period,name={Table}}
\raggedbottom
\renewcommand{\thetable}{\arabic{table}}

\title{Hierarchy-Boosted Funnel Learning for Identifying Semiconductors with Ultralow Lattice Thermal Conductivity}

\author{Mengfan Wu}
\affiliation{Shanghai Research Institute for Intelligent Autonomous Systems, Tongji University, Shanghai 200092, China}
\affiliation{Center for Phononics and Thermal Energy Science, Shanghai Key Laboratory of Special Artificial Microstructure Materials and Technology, School of Physics Science and Engineering, Tongji University, Shanghai 200092, China}

\author{Shenshen Yan}
\affiliation{Center for Phononics and Thermal Energy Science, Shanghai Key Laboratory of Special Artificial Microstructure Materials and Technology, School of Physics Science and Engineering, Tongji University, Shanghai 200092, China}

\author{Jie Ren}
\email[Corresponding author\\E-mail: ]{Xonics@tongji.edu.cn}
\affiliation{Shanghai Research Institute for Intelligent Autonomous Systems, Tongji University, Shanghai 200092, China}
\affiliation{Center for Phononics and Thermal Energy Science, Shanghai Key Laboratory of Special Artificial Microstructure Materials and Technology, School of Physics Science and Engineering, Tongji University, Shanghai 200092, China}

\date{\today}
\begin{abstract}
Data-driven machine learning (ML) has demonstrated tremendous potential in material property predictions. However, the scarcity of materials data with costly property labels in the vast chemical space presents a significant challenge for ML in efficiently predicting properties and uncovering structure-property relationships. Here, we propose a novel hierarchy-boosted funnel learning (HiBoFL) framework, which is successfully applied to identify semiconductors with ultralow lattice thermal conductivity ($\kappa_\mathrm{L}$). By training on only a few hundred materials targeted by unsupervised learning from a pool of hundreds of thousands, we achieve efficient and interpretable supervised predictions of ultralow $\kappa_\mathrm{L}$, thereby circumventing large-scale brute-force \textit{ab initio} calculations without clear objectives. As a result, we provide a list of candidates with ultralow $\kappa_\mathrm{L}$ for potential thermoelectric applications and discover a new factor that significantly influences structural anharmonicity. This HiBoFL framework offers a novel practical pathway for accelerating the discovery of functional materials.
\end{abstract}

\maketitle

\section{Introduction}
Emerging as a powerful technology of the data-driven paradigm in materials science, machine learning (ML) has considerably accelerated the design and discovery of promising materials in recent years,\cite{mueller2016machine, choudhary2022recent, luo2023predicting, chen2022uv, fan2025combining} including ML interatomic potentials,\cite{deringer2019machine} inverse design of materials,\cite{yao2021inverse} efficient property predictions.\cite{wu2023target, wang2024interpretable} Simultaneously, the advancements in high-performance computing greatly facilitate the establishment of diverse material-related databases utilizing density functional theory (DFT)-based high-throughput calculations (HTC), such as the Materials Project (MP),\cite{jain2013commentary} Open Quantum Materials Database (OQMD),\cite{saal2013materials, kirklin2015open} Automatic-FLOW for Materials Discovery (AFLOW),\cite{curtarolo2012aflow} Joint Automated Repository for Various Integrated Simulations (JARVIS),\cite{choudhary2020joint} etc. These continuously expanding databases also lay a solid foundation for the application of cutting-edge ML technology in the field of materials science. 

As the two main categories of ML, namely supervised and unsupervised learning strategies, both have achieved remarkable success in different ways. On the one hand, supervised learning enables the efficient predictions of material properties by-passing solving the complex equations of quantum mechanics based on expensive DFT calculations, which emphasizes the requirement of large human-labelled datasets for model training to ensure the accuracy. Ridge regression,\cite{stuke2019chemical} decision tree,\cite{takahashi2019creating} support vector machine,\cite{oliynyk2016classifying} random forest,\cite{torrisi2020random} gradient boosting decision tree,\cite{ren2024machine} etc., are widely employed in designing and screening potential materials with desired properties, which also elucidate the close relationship between the structure and target property.\cite{bharadwaj2024unlocking, xu2024universal, zhang2024synthesis} On the other hand, as a technology operates without the necessity of well-labeled training data, unsupervised learning possesses the capability to infer the underlying patterns among varieties of materials within a feature space. Predominantly employed methods in unsupervised learning encompass clustering, dimensionality reduction, and anomaly detection, in which clustering can categorize different materials into the corresponding clusters by assessing their similarities between each other, thereby identifying candidates resembling the anticipated data points.\cite{zou2023unsupervised, jia2022unsupervised, wang2021unsupervised} In this context, combining HTC and ML technologies can not only efficiently explore novel materials in the vast chemical space but also gain insights into the structure-property relationship at a quantitative level. However, a significant challenge lies in labeling data for materials with intrinsically complex properties in the vast chemical space, especially the lattice thermal conductivity ($\kappa_\mathrm{L}$), due to the complexities involved in experimental measurements and the unbearable computational costs associated with accurate DFT calculations. Is there an effective approach to reduce the cost of labeling, while enabling the efficient prediction of complex properties and elucidation of structure-property relationships?

\begin{figure*}[!t]
\centering
\includegraphics[width=1\textwidth]{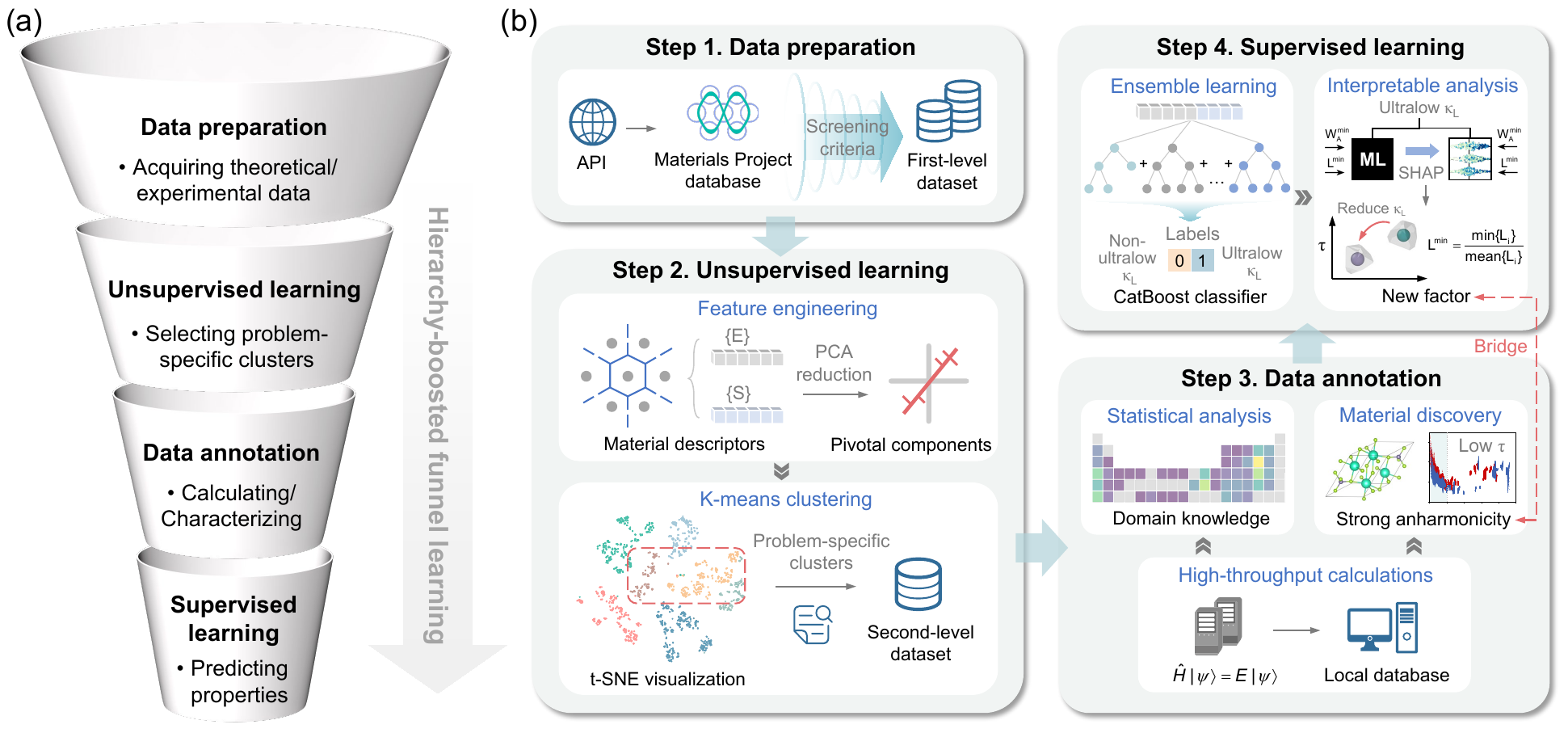}
\caption{(a) Schematic of the HiBoFL framework, including data preparation, unsupervised learning, data annotation and supervised learning. (b) Workflow of applying the HiBoFL framework to efficiently identify semiconductors with ultralow $\kappa_\mathrm{L}$.}
\label{figl}
\end{figure*}

Functional materials exhibiting ultralow $\kappa_\mathrm{L}$ possess vital significance across various fields, such as power generation,\cite{sootsman2009new} heat conduction,\cite{pernot2010precise} thermal barrier coatings\cite{padture2002thermal} and so on,\cite{gao2018unusually, wang2021strain, xiang2022thermal} which greatly advance the development of industry. Particularly, owing to the key role in directly converting heat energy into electricity based on the thermoelectric (TE) effect, TE materials have become the focal point of considerable interest in academic and industrial research. Indeed, the conversion efficiency of a TE material is theoretically quantified by its figure of merit $zT$: $zT={S^{2}\sigma}T/{\kappa}$, where $S$ is the Seebeck coefficient, $\sigma$ is the electric conductivity, $\kappa$ is the thermal conductivity, and $T$ is the absolute temperature, respectively. Moreover, $\kappa$ can be expressed in two parts: $\kappa=\kappa_\mathrm{e}+\kappa_\mathrm{L}$, where $\kappa_\mathrm{e}$ and $\kappa_\mathrm{L}$ are the electronic thermal conductivity and lattice thermal conductivity, respectively, indicating that they all contribute to heat conduction. For metallic materials, $\kappa_\mathrm{e}$ plays a dominant role in heat conduction due to the presence of a large number of free electrons. On the contrary, in semiconductors or insulators, thermal energy is predominantly transferred through lattice vibrations, with $\kappa_\mathrm{L}$ making the primary contribution. In this process, the quanta of such lattice vibrations in a solid are the so-called phonons.

As clearly noticed, toward the goal of seeking TE materials with optimal $zT$ values, it requires not only maximizing the power factor ($S^{2}\sigma$) but also minimizing the thermal conductivity ($\kappa_\mathrm{e}+\kappa_\mathrm{L}$) simultaneously.\cite{snyder2008complex} Notably, the distinct and separate scale of the mean free paths for electrons and phonons contributes to the independence of $\kappa_\mathrm{L}$ as a parameter in $zT$, which makes decreasing $\kappa_\mathrm{L}$ become a significant avenue to realize the ideal concept of “phonon-glass and electron-crystal”,\cite{nolas1999skutterudites, snyder2004disordered} ultimately leading to excellent TE performance. For the past decades, unremitting efforts have been made to discover a series of materials with $\kappa_\mathrm{L}$, for instance, Tl\textsubscript{3}VSe\textsubscript{4},\cite{mukhopadhyay2018two} TlInTe\textsubscript{2},\cite{jana2017intrinsic} InTe,\cite{jana2016origin} CsAg\textsubscript{5}Te\textsubscript{3},\cite{lin2016concerted} AgSbSe\textsubscript{2},\cite{nielsen2013lone} etc. However, the conversion efficiency of TE materials has persistently been a significant challenge in the efficient recovery of waste heat, that is to say, useful materials featuring ultralow $\kappa_\mathrm{L}$ are still in urgent demand. Despite the few successes of traditional trial-and-error experiments and case-by-case DFT calculations in the exploration of desired materials, efficient and robust material design oriented towards $\kappa_\mathrm{L}$ in the vast chemical space is hindered by the complex structure-property relationship and size-limited resources.

In the present work, we propose a novel hierarchy-boosted funnel learning (HiBoFL) framework that integrates unsupervised learning and supervised learning to efficiently achieve complex property predictions, which is applied to identify semiconductors with ultralow $\kappa_\mathrm{L}$. Unsupervised learning is used to uncover underlying patterns among different materials, facilitating the identification of specific clusters with a high likelihood of exhibiting ultralow $\kappa_\mathrm{L}$. Based the low-cost HTC on this significantly reducing space, we establish a local database and discover a series of semiconductors with ultralow $\kappa_\mathrm{L}$, in which Cs\textsubscript{2}SnSe\textsubscript{3} and Cs\textsubscript{2}GeSe\textsubscript{3} are screened out for in-depth mechanism analysis. Furthermore, a supervised classification model for directly predicting ultralow $\kappa_\mathrm{L}$ is trained to refine the results. With resolved important descriptors that govern ultralow $\kappa_\mathrm{L}$, we are capable of investigating the $\kappa_\mathrm{L}$ modulation mechanism and uncovering a new factor that governs structural anharmonicity. We expect that this HiBoFL framework can also be widely applied in the discovery of other functional materials with excellent performances.

\section{Results and Discussion}
\subsection{HiBoFL Framework for Accelerating the Discovery of Functional Materials}
Our proposed novel HiBoFL framework for accelerating the discovery of functional materials with complex properties is shown in {\color{blue}{Figure 1a}}, which exhibits a funnel-like structure driven by a hierarchical framework, effectively narrowing the search space while boosting model performances. This framework mainly includes four parts: I) Data preparation. An initial theoretical or experimental dataset of the target material system is required, potentially involving preliminary high-throughput screening, data cleaning, and other preprocessing operations. II) Unsupervised learning. Relevant features are extracted from the initial dataset, encompassing aspects such as experimental process parameters, intrinsic material properties, etc. By employing clustering algorithms, distinct classes of data points with potentially similar properties can be identified, in which problem-specific clusters are selected to narrow the search space. III) Data annotation. Data from the problem-specific clusters can be assigned corresponding property labels through further relatively low-cost experimental characterizations or HTC. This facilitates the establishment of a local database, enabling the extraction of prior domain knowledge through statistical data analysis. IV) Supervised learning. The labeled dataset within the established local database can be used to further train supervised learning models, which helps refine the coarse results from unsupervised learning, ultimately enabling direct and rapid prediction of the target properties. Such a HiBoFL framework can not only reduce the expensive cost of labeling data, but also efficiently predict the costly-labeled complex properties.

We then apply this HiBoFL framework to efficiently identify semiconductors with ultralow $\kappa_\mathrm{L}$ as shown in {\color{blue}{Figure 1b}}. 
In the first step, we obtain the material dataset from the MP database based on its application programming interface (API) and then apply a series of screening criteria to derive the first-level dataset for subsequent research. 
In the second step, chemical composition descriptors based on Magpie\cite{ward2016general} and crystal structure descriptors derived from Voronoi tessellations\cite{ward2017including} are used to featurize the materials in the first-level dataset. We use principal component analysis (PCA),\cite{abdi2010principal} for dimensionality reduction thereby obtaining pivotal components. \textit{K}-means clustering\cite{hartigan1979algorithm, likas2003global} is then employed to identify materials with similar $\kappa_\mathrm{L}$, which is visualized in a low-dimensional space using t-distributed stochastic neighbor embedding (t-SNE),\cite{van2008visualizing} and problem-specific clusters are selected based on similarity design rules to form the second-level dataset. 
In the third step, we use the phonon-elasticity-thermal (PET) model\cite{yan2022high} to perform low-cost HTC on the materials in the second-level dataset, establishing a local database based on the HTC results. On the one hand, a list of candidates for potential TE applications can be recommended and their statistical analysis can help us to summarize several domain knowledge of ultralow $\kappa_\mathrm{L}$. On the other hand, we can directly screen out candidate materials with ultralow $\kappa_\mathrm{L}$ for in-depth mechanism analysis, where the results are verified by accurately solving the phonon Boltzmann transport equation (BTE) and the phonon thermal transport mechanisms are further revealed. 
In the fourth step, we perform ensemble learning on the labeled local database for directly classifying ultralow $\kappa_\mathrm{L}$, training a robust CatBoost classifier\cite{prokhorenkova2018catboost}. The interpretable analysis of the pre-trained model based on the SHapley Additive exPlanations (SHAP) method\cite{shapley1953value, lundberg2017adv, lundberg2020local} further reveals the influence of important descriptors on ultralow $\kappa_\mathrm{L}$. A new factor capable of significantly reducing $\kappa_\mathrm{L}$ via enhancing structural anharmonicity is discovered, eventually building a bridge between the ML model interpretability and first-principles analysis. 

\begin{figure}[!t]
\centering
\includegraphics[width=0.4\textwidth]{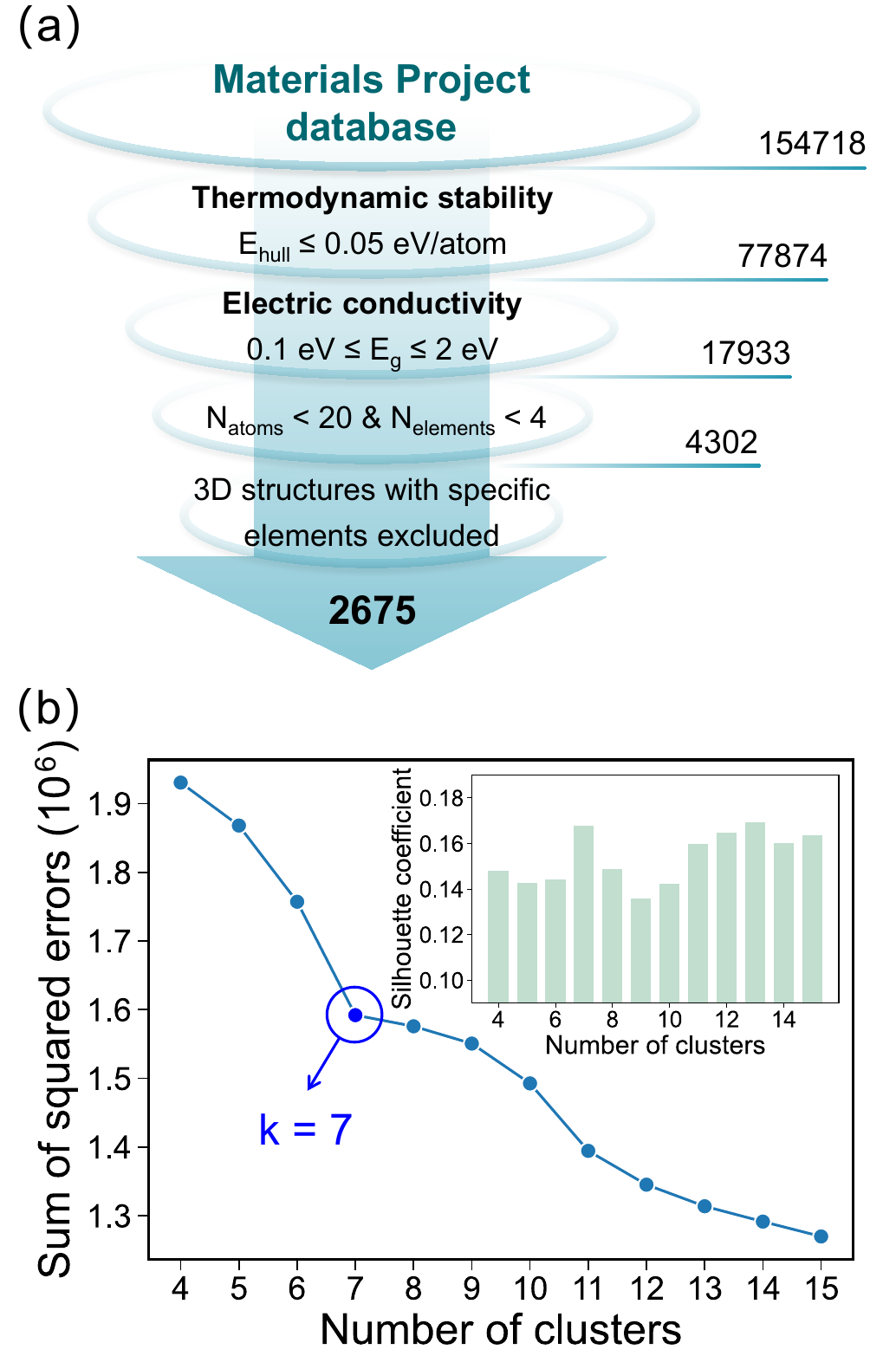}
\caption{(a) Flowchart of preliminary high-throughput screening from the Materials Project (MP) database. (b) Optimization of the number of clusters \textit{k} for the \textit{k}-means algorithm based on the elbow method and silhouette coefficient (inset).}
\label{figl}
\end{figure}

\subsection{Preliminary High-Throughput Screening}
The process of preliminary high-throughput screening to select materials for subsequent ML investigations is illustrated in {\color{blue}{Figure 2a}}. Initially, we start by acquiring all the materials from the MP database based on its API, resulting in a total of 154718 entries saved in a JSON file as the Python dictionary object. Of these, several specific criteria are set to narrow down the range of exploration. The first screening criterion focuses on the assessment of thermodynamic stability, in which the materials with energy above the convex hull ($E_\mathrm{hull}$, the formation energy difference between the target compound and its competing phases) no more than 0.05 eV/atom are considered to show a high likelihood of being synthesized in experiments. Further, the band gap ($E_\mathrm{g}$) can be directly retrieved from the MP database, set within the range of 0.1–2 eV for assessing electric conductivity. This specific interval is highly characteristic of semiconductors, demonstrating the inherent capacity of these screened materials for favorable electrical conductivity. Additionally, taking into account of the computational cost associated with $\kappa_\mathrm{L}$, our constraints on the material system involve ensuring that the number of atoms ($N_\mathrm{atoms}$) is less than 20 in the unit cell of a crystal structure and the number of elements ($N_\mathrm{elements}$) is below four in one compound, respectively. Ultimately, through excluding the materials containing hydrogen, lanthanides, and actinides, and conducting structural analysis, we further obtain 2675 three-dimensional (3D) crystal structures without calculation errors. These materials constitute the first-level dataset for our in-depth investigations.
    
\subsection{Unsupervised Learning for Identifying Materials with  Similar $\kappa_\mathrm{L}$}
\begin{figure*}[htbp]
\centering
\includegraphics[width=1\textwidth]{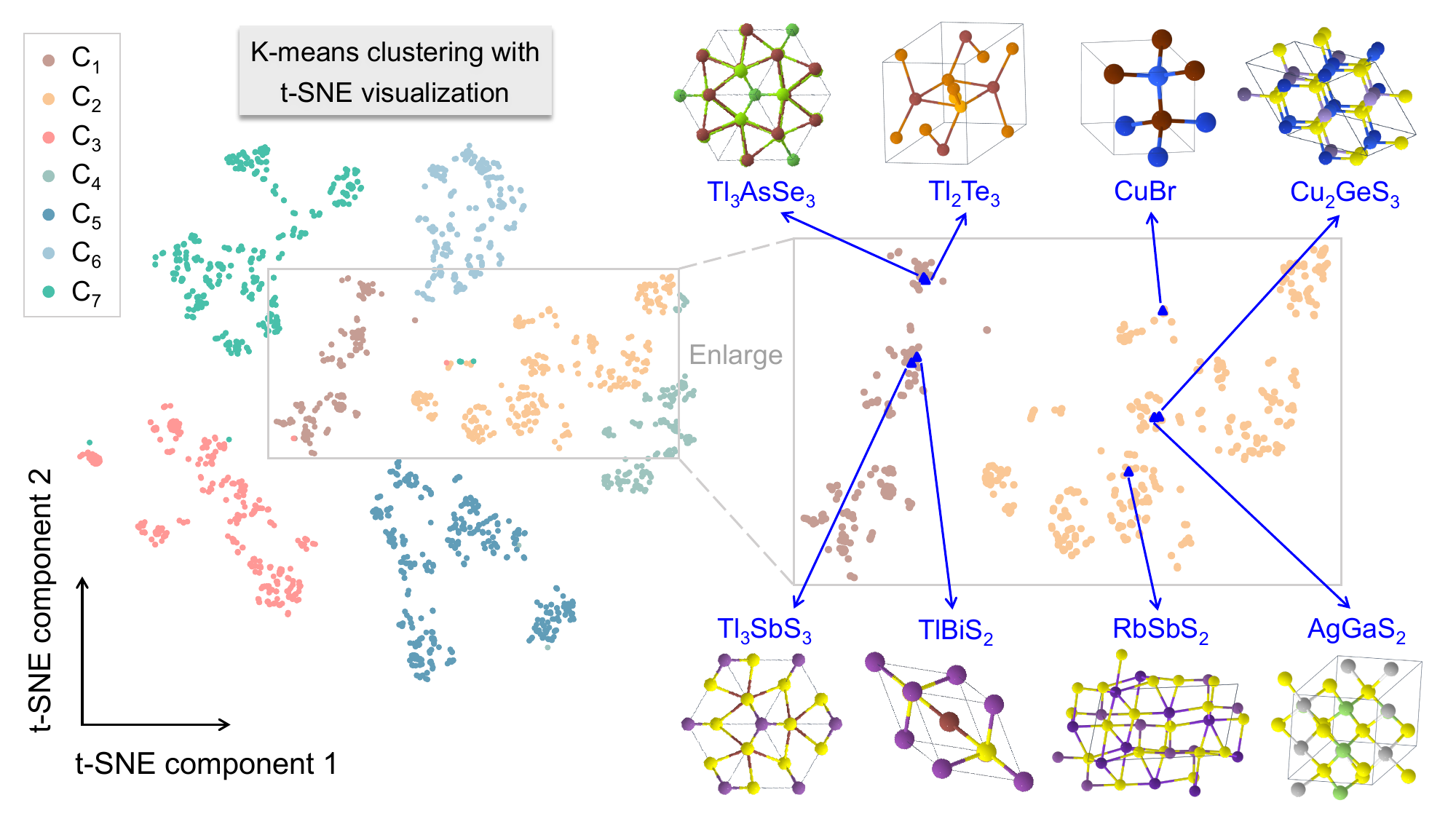}
\caption{Unsupervised learning of the materials in the first-level dataset.  Using t-SNE visualization for the seven clusters generated by the \textit{k}-means algorithm, where each point represents a compound and is colored with the corresponding cluster. Eight represented materials in \textit{C}\textsubscript{1} and \textit{C}\textsubscript{2} with low experiment-measured $\kappa_\mathrm{L}$ are marked in the enlarged region.}
\label{Figure 3}
\end{figure*}

We next carry out unsupervised learning to identify materials with a high likelihood of exhibiting relatively low $\kappa_\mathrm{L}$ based on similarity design rules. Following the generation of the first-level dataset as input, the materials should be transformed into the length-fixed vector as the so-called descriptors, which are the key point to distinguish different materials. Herein, we use two distinct types of descriptors to featurize these materials, namely, composition-based features and structure-based features, to account for the mapping relationships between different categories of descriptors and \textit{$\kappa_\mathrm{L}$} from the perspectives of chemical composition and crystal structure. Among them, composition-based features are generated based on Magpie data and denoted by \{E\}, including electronegativity, atomic number, fraction of electrons, etc., which are comprehensive enough to capture the characteristics of chemical compositions with different constituent elements and proportions. Through partitioning the crystal structures into Wigner-Seitz cells of each atom, a series of structure-based features can be derived from Voronoi tessellations based on the characteristics of the local environment of each atom in the unit cell, which are denoted by \{S\}. In this manner, each material can generate a total of 273 descriptors automatically without any time-consuming DFT calculations based on its corresponding composition and structure, jointly denoted by \{E, S\}. Notably, these descriptors have been successfully applied to the ML predictions of $\kappa_\mathrm{L}$ at various temperatures, indicating their significant potential for mapping the property of $\kappa_\mathrm{L}$. After feature generation, it is crucial to preprocess all the data thereby enhancing the performance of clustering model. To address the large variations among different feature values and transform them to follow a normal distribution, both standardization and quantile transformer are employed for preprocessing these descriptors. Further, we perform PCA based on singular value decomposition to project the preprocessed features to a lower dimensional space. PCA linearly combines original features into principal components (PCs), ensuring the hierarchical order based on their contributions, in which the first principal component captures the largest explainable variance, followed by the second one and so forth. From the curve of total explainable variance changing with the number of PCs as shown in {\color{blue}{Figure S1}}, we can conclude that 83 PCs suffice to account for 99\% of the variance among all 2675 materials in the first-level dataset. Hence, these 83 PCs are extracted as the pivotal components for input into the following clustering algorithm. 

Subsequently, the \textit{k}-means algorithm is utilized to identify the underlying associations with $\kappa_\mathrm{L}$ from these unlabeled materials, in which only one critical parameter is required to be predefined, i.e., the number of clusters \textit{k}.\cite{kodinariya2013review} Based on the analysis of the elbow method\cite{cui2020introduction} and the silhouette coefficient\cite{shahapure2020cluster} as depicted in {\color{blue}{Figure 2b}}, the elbow-like point of inflection on the curve emerges at a value of \textit{k} = 7, indicating the location where inertia or distortion starts decreasing significantly at a very slow rate, represents the optimal \textit{k} value for the \textit{k}-means clustering. At this stage, a relatively high silhouette coefficient also illustrates that each material exhibits strong similarity within its respective cluster, while different clusters are well-separated as much as possible. As a result, we partition the first-level dataset into seven clusters (from \textit{C}\textsubscript{1}, \textit{C}\textsubscript{2}, ..., to \textit{C}\textsubscript{7}), on the basis of similarities of these feature vectors derived from chemical compositions and crystal structures, in which the materials in the same cluster are considered to show a high likelihood of possessing similar $\kappa_\mathrm{L}$. Since the first two PCs only capture about 35\% of the variance within entire data, it is insufficient for the clustering results to be intuitively visualized in a two-dimensional (2D) mapping. Unlike PCA, which emphasizes preserving large pairwise distances to maximize variance, t-SNE is a powerful non-linear dimensionality reduction technology for the visualization of high-dimensional data, aiming to maintain pairwise similarities among data points in a lower-dimensional space. We strictly use t-SNE only for the visualization of the resulting seven clusters of \textit{k}-means algorithm as shown in {\color{blue}{Figure 3}}, projecting the 83 PCs into a 2D latent space comprised of two t-SNE components. The result intuitively demonstrates the high-quality clustering with distinct separation between each cluster, where each point represents a material and its color relates to the corresponding category of cluster.

Since we have obtained seven clusters through \textit{k}-means clustering, it is necessary to conduct the similarity analysis of these clusters. Materials within the same cluster are considered to show similar structures thereby likely sharing similar properties, which facilitates a deep comprehension of underlying patterns and relationships among the 2675 materials. To evaluate the similarity criteria of these materials, the reported $\kappa_\mathrm{L}$ values at 300 K for several known materials included in each cluster are collected from the previous studies (most are experimentally measured), which are all listed in {\color{blue}{Table S1}} along with some other basic information. As we expected, materials within the same cluster exhibit closely similar $\kappa_\mathrm{L}$ values, while there is a comparatively significant difference in $\kappa_\mathrm{L}$ among materials from different clusters. Particularly, the known materials with relatively low $\kappa_\mathrm{L}$ are clustered into \textit{C}\textsubscript{1} and \textit{C}\textsubscript{2} through \textit{k}-means clustering, including eight structures of Tl\textsubscript{3}AsSe\textsubscript{3} (0.23 W/mK),\cite{ewbank1982thermal} Tl\textsubscript{2}Te\textsubscript{3} (0.40 W/mK),\cite{spitzer1970lattice} Tl\textsubscript{3}SbS\textsubscript{3} (0.42 W/mK),\cite{skoug2011role} TlBiS\textsubscript{2} (0.80 W/mK),\cite{popovich2003electrical} CuBr (1.30 W/mK),\cite{haynes2016crc} Cu\textsubscript{2}GeS\textsubscript{3} (1.20 W/mK),\cite{spitzer1970lattice} RbSbS\textsubscript{2} (1.60 W/mK)\cite{skoug2011role} and AgGaS\textsubscript{2} (1.50 W/mK),\cite{beasley1994thermal} which are all clearly shown in the enlarged part of {\color{blue}{Figure 3}}. On the contrary, \textit{C}\textsubscript{7} contains materials with apparently large $\kappa_\mathrm{L}$, including four structures of GaN (130 W/mK),\cite{slack1987intrinsic} BP (350 W/mK),\cite{morelli2006high, slack1973nonmetallic} SiC (490 W/mK)\cite{shinde2006high} and Si (156 W/mK).\cite{glassbrenner1964thermal} Thus, as confirmed by the good distinction between low and high $\kappa_\mathrm{L}$ among each cluster, our proposed unsupervised learning model demonstrates great potential to identify compositional and structural information about the $\kappa_\mathrm{L}$ of these materials, leading to the successful clustering into different categories according to this property. Since the materials with a high likelihood of possessing relatively low $\kappa_\mathrm{L}$ tend to group into the two clusters of \textit{C}\textsubscript{1} and \textit{C}\textsubscript{2}, the exploration scope for finding materials with low $\kappa_\mathrm{L}$ is reduced from 2675 materials to 704 materials, narrowing by approximately three-quarters. As a result, these two problem-specific clusters (\textit{C}\textsubscript{1} and \textit{C}\textsubscript{2}) further constitute our second-level dataset. 
\begin{figure*}
\centering
\includegraphics[width=1\textwidth]{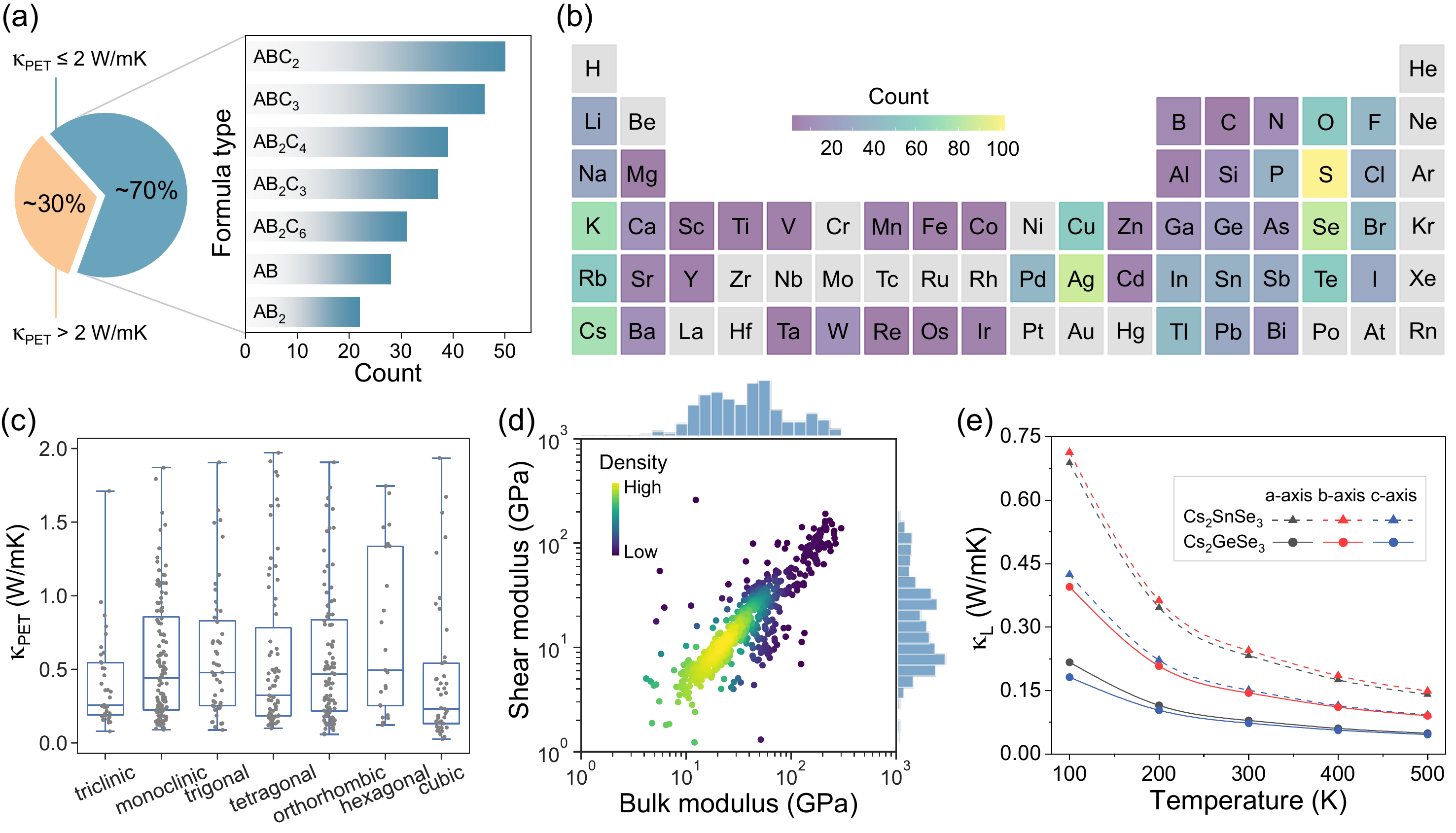}
\caption{ (a–d) Statistical analysis of HTC results in the second-level dataset. (a) Pie chart of $\kappa_\mathrm{PET}$ separated by a threshold of 2 W/mK and distribution of different formula types with counts exceeding 20. (b–d) Distribution of the materials with $\kappa_\mathrm{PET}$ no greater than 2 W/mK: (b) Heat map over the periodic table of elements for easy visualization of element counts. (c) Box plot of different crystal systems. (d) Density scatter plot of  shear modulus and bulk modulus. (e) Calculated $\kappa_\mathrm{L}$ as the function of temperature at different axes in Cs\textsubscript{2}SnSe\textsubscript{3} and Cs\textsubscript{2}GeSe\textsubscript{3} by solving the phonon BTE.} 
\label{figl}
\end{figure*}

\subsection{HTC of Specific Clusters and Statistical Analysis }
Given the success of unsupervised learning in significantly reducing the broad material search space, it has become feasible to further extend the second-level dataset into a labeled repository of thermal conductivity through HTC at affordable computational costs. Based on the PET empirical equation within the high-throughput framework, we derive the $\kappa_\mathrm{PET}$ values at 300 K ignoring the anisotropy for materials in the second-level dataset, while excluding structures that do not satisfy the mechanical stability criteria. The basic information and corresponding $\kappa_\mathrm{PET}$ values of the resulting 661 materials are then stored in a local database using MongoDB,\cite{MongoDB} serving as a valuable repository to retrieve data and prioritize detailed theoretical and experimental investigations. Of particular note is that nearly 70\% of these semiconductors exhibit the $\kappa_\mathrm{PET}$ values no greater than 2 W/mK ({\color{blue}{Figure 4a}}), with a considerable portion of these materials remaining unreported to date. These materials also constitute a list of candidate materials with potential applications in the TE field ({\color{blue}{Table S2}}), which undoubtedly prove that the specific clusters we identified through unsupervised learning indeed contain a significant number of materials with low thermal conductivity.  

In addition, we also count the distribution of different formula types with counts exceeding 20 in these materials. Apparently, the structures represented by the two types of formula anonymous dominate in quantity, namely the ABC\textsubscript{2} type characterized by the diamond-like structure, and the ABC\textsubscript{3} type represented by the perovskite-type structure. To gain a more intuitive insight into the distribution patterns of materials with low thermal conductivity, we plot a heat map over the periodic table of elements in {\color{blue}{Figure 4b}}, illustrating the count of elements present in these compounds with $\kappa_\mathrm{PET}$ no greater than 2 W/mK. Sulfur, selenium, tellurium, and oxygen belonging to the chalcogens consecutively occupy the largest counts among the anion elements, in which sulfur is the most abundant with a count of 102. As for the cation elements, silver, cesium, potassium, copper, and rubidium occupy the highest abundance, respectively. This can be explained by the fact that materials composed of heavy elements or characterized by weak chemical bonding typically exhibit lower $\kappa_\mathrm{L}$.
Notably, a few previous studies have constrained the search space of materials within the above-mentioned characteristics to investigate those with low thermal conductivity, such as high-throughput screening in chalcogenide ABC\textsubscript{3} perovskites\cite{cao2024high} or diamond-like ABC\textsubscript{2} compounds,\cite{li2019high} detailed analysis in the IV–VI chalcogenides\cite{guillemot2024impact} and so forth.\cite{deng2021electronic, plata2022charting, posligua2023theoretical, nielsen2013lone}
The box plot of different crystal systems indicates that the $\kappa_\mathrm{PET}$ values of materials in each crystal system primarily cluster around 0.5 W/mK, with the monoclinic system exhibiting the largest quantity ({\color{blue}{Figure 4c}}). We also show the density distributions of the shear modulus and bulk modulus obtained by the Voigt–Reuss–Hill (VRH) method as shown in {\color{blue}{Figure 4d}}, which are primarily concentrated within the relatively small range of $ \sim $50 GPa. This suggests that materials with lower shear and bulk moduli might be more likely to exhibit lower thermal conductivity. These statistical analyses provide deep insights into the regulation of $\kappa_\mathrm{L}$, which can offer prior domain knowledge for researchers in selecting specific systems to obtain ultralow $\kappa_\mathrm{L}$.

To further validate the results and conduct in-depth analysis of the phonon thermal transport mechanism, we calculate more precise $\kappa_\mathrm{L}$ values according to first-principles derived force constants and Boltzmann transport theory for the unreported materials Cs\textsubscript{2}SnSe\textsubscript{3} and Cs\textsubscript{2}GeSe\textsubscript{3}, which are screened out based on the formula type ranking among the materials with the lowest $\kappa_\mathrm{PET}$ values. {\color{blue}{Figure 4e}} depicts the calculated $\kappa_\mathrm{L}$ as the function of temperature ranging from 100 to 500 K at different axes in the discussed semiconductors, in which all the intrinsic $\kappa_\mathrm{L}$ values show obvious anisotropy and gradually decrease following the T\textsuperscript{–1} manner with the temperature increasing just as the hallmark of Umklapp scattering. The rise in temperature results in an elevation in the equilibrium phonon population, consequently leading to intense phonon-phonon collisions, as delineated by $\overline{n}\approx{k_{\mathrm{B}}T}/{\hbar\omega}\ (T\gg\Theta_{\mathrm{D}})$, where $\overline{n}$ is the average number of phonons, $k_{\mathrm{B}}$ is the Boltzmann constant, \textit{T} is the temperature, $\omega$ is the phonon frequency, $\hbar$ is the reduced Planck constant and $\Theta_{\mathrm{D}}$ is the Debye temperature. Moreover, these materials all exhibit intrinsically ultralow $\kappa_\mathrm{L}$, with values all below 0.25 W/mK in any direction at 300 K, significantly lower than the $\kappa_\mathrm{L}$ ($ \sim $2.3 W/mK) of the traditional TE material PbTe.\cite{pei2011convergence} The results indicate the potential application of Cs\textsubscript{2}SnSe\textsubscript{3} and Cs\textsubscript{2}GeSe\textsubscript{3} in the TE field, further substantiating the effectiveness of our previously adopted approach—combining unsupervised learning with HTC to discover semiconductors with ultralow $\kappa_\mathrm{L}$. 

\begin{figure*}
\centering
\includegraphics[width=1\textwidth]{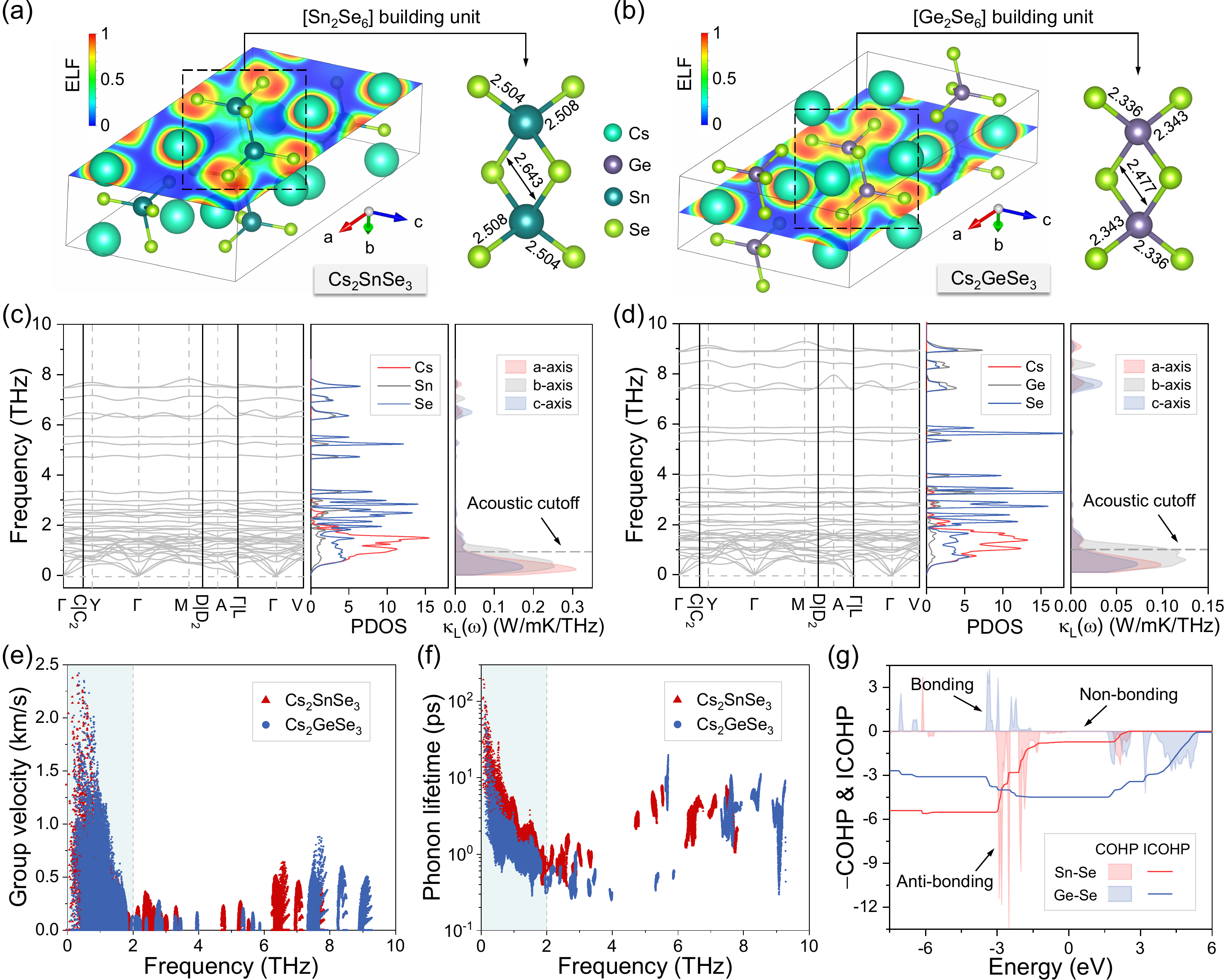}
\caption{(a–g) First-principles analysis of the phonon thermal transport properties. Crystal structures and the projected 2D ELF diagram of (a) Cs\textsubscript{2}SnSe\textsubscript{3} and (b) Cs\textsubscript{2}GeSe\textsubscript{3}. Phonon dispersion (left panels), atom-projected PDOS (middle panels) and spectral $\kappa_\mathrm{L}(\omega)$ (right panels) of (c) Cs\textsubscript{2}SnSe\textsubscript{3} and (d) Cs\textsubscript{2}GeSe\textsubscript{3}. (e) Group velocity along the \textit{b}-axis and (f) phonon lifetime as a function of frequency at 300 K for Cs\textsubscript{2}SnSe\textsubscript{3} and Cs\textsubscript{2}GeSe\textsubscript{3}. (g) COHP and ICOHP projected on Sn–Se and Ge–Se bonds.}
\label{figl}
\end{figure*}

\subsection{Mechanisms of Phonon Thermal Transport Properties}
Generally speaking, compounds with heavier atoms tend to exhibit lower group velocities due to the reduced phonon frequency. As a result, a lower $\kappa_\mathrm{L}$ of Cs\textsubscript{2}SnSe\textsubscript{3} was expected given the relatively heavy nature of Sn. However, it is noteworthy that the $\kappa_\mathrm{L}$ of Cs\textsubscript{2}SnSe\textsubscript{3} is obviously higher than that of Cs\textsubscript{2}GeSe\textsubscript{3} (i.e., $ \sim $3 times in the \textit{a}-axis, $ \sim $1.7 times in the \textit{b}-axis, $ \sim $2 times in the \textit{c}-axis at 300 K), which presents an interesting unusual phenomenon. Next, we would like to discuss the microscopic mechanisms responsible for the ultralow $\kappa_\mathrm{L}$ and unusual difference observed among the two materials.

Cs\textsubscript{2}SnSe\textsubscript{3} and Cs\textsubscript{2}GeSe\textsubscript{3} both crystallize in the same space group \textit{C}2/\textit{m} (No. 12) of the monoclinic crystal system, with the fully optimized crystallographic parameters listed in {\color{blue}{Table S3}}. Their crystal structures are shown in {\color{blue}{Figures 5a and 5b}}. 
Sn/Ge atoms are coordinated by four Se atoms in a tetrahedral geometry [SnSe\textsubscript{4}]/[GeSe\textsubscript{4}] at distances in 2.504–2.643 \text {Å}/2.336–2.477 \text {Å}, and these tetrahedra further share a common edge to form dimeric [Sn\textsubscript{2}Se\textsubscript{6}]/[Ge\textsubscript{2}Se\textsubscript{6}] building units, respectively, which form the four-membered [Sn\textsubscript{2}Se\textsubscript{2}]/[Ge\textsubscript{2}Se\textsubscript{2}] rings and are charge compensated by Cs cations.
The phase diagrams for Cs–Sn–Se and Cs–Ge–Se systems based on the calculated energies in the MP database show that Cs\textsubscript{2}SnSe\textsubscript{3} and Cs\textsubscript{2}GeSe\textsubscript{3} possess similar thermodynamical stability with equal (zero) convex hull distances ({\color{blue}{Figure S2}}). Simultaneously, the phonon dispersion curves along high-symmetry directions in Brillouin zone indicate that they are dynamically stable due to the absence of imaginary phonon modes (left panels in {\color{blue}{Figures 5c and 5d}}). Thus, the results demonstrate the feasibility of experimentally synthesizing these two materials. 
To identify the bonding characteristics in Cs\textsubscript{2}SnSe\textsubscript{3} and Cs\textsubscript{2}GeSe\textsubscript{3}, we employ the electron localization function (ELF) to quantify the extent of spatial localization of the reference electron with values ranging from 0 to 1.
The localization of electrons in the Sn/Ge–Se bonding region illustrates the covalent nature of the Sn/Ge–Se bonds, in which a polar covalent bond between Sn and Se (ELF $\approx{0.5}$) is observed due to the smaller electronegativity and larger atomic radius of Sn, leading to a significant difference among the two compounds. On the contrary, there is no overlapping of charge clouds between Cs atoms and other atoms, indicating the presence of strong ionic bonding which aligns with the fact that a relatively large electronegativity (on the pauling scale) difference (1.76) between Cs (0.79) and Se (2.55) results in strong ionic characteristics. Therefore, Cs\textsubscript{2}SnSe\textsubscript{3} and Cs\textsubscript{2}GeSe\textsubscript{3} contain multiple types of bonds, i.e., ionic Cs–Se and covalent Sn/Ge–Se, resulting in complex crystal structures with bonding hierarchy.
These complex structures, comprising heavy atoms, weakly bound and rigid distorted units with a significant bonding hierarchy, are anticipated to exhibit large lattice anharmonicity.\cite{heremans2015anharmonicity}

The phonon dispersion reveals that there are 3 acoustic branches and 33 optical branches at each phonon wave vector \textit{q} due to the 12 atoms in the primitive cell of both Cs\textsubscript{2}SnSe\textsubscript{3} and Cs\textsubscript{2}GeSe\textsubscript{3}, in which three acoustic branches are composed of one longitudinal acoustic mode (LA) and two transverse acoustic modes (TA and TA'). A striking common feature of their phonon dispersion is that a waterfall-like low-lying optical branch (LLO) exhibits the avoided crossing behavior with acoustic modes around the Brillouin zone center, resulting in strong acoustic-optical coupling as one of the potential signals of the rattling model.\cite{zeng2022physical} These characteristics can not only lead to a softening of the acoustic modes thereby yielding low phonon group velocities, but also greatly enhance the scattering rates of heat-carrying acoustic phonons, all of which contribute to the suppression of $\kappa_\mathrm{L}$ for Cs\textsubscript{2}SnSe\textsubscript{3} and Cs\textsubscript{2}GeSe\textsubscript{3}. 
Analysis of the atom-projected PDOS and spectral $\kappa_\mathrm{L}(\omega)$ (middle and right panels in {\color{blue}{Figures 5c and 5d}}) indicate that acoustic and LLO phonon modes in the low-frequency range (0–2 THz) are primarily dictated by Cs and Se vibrations followed by Sn/Ge vibrations, which also make predominant contributions to the ultralow $\kappa_\mathrm{L}$. A localized region mostly contributed by Cs atoms within a narrow energy window centered around 1 THz is observed in both compounds, indicating the anharmonic rattling-like motion of the weakly bonded Cs atoms, which is responsible for the presence of the soft LLO modes. 
This can also be confirmed in the potential energy curves obtained by shifting the atoms with respect to their static equilibrium positions along different axes as depicted in {\color{blue}{Figure S3}}. Sn/Ge and Se atoms are both confined within the comparatively steep potential wells, whereas Cs atoms with heavier mass can vibrate easily with larger amplitude due to the shallowest potential energy surface in all the directions. These atoms exhibit nearly identical displacement magnitudes in their respective compounds. Therefore, the loosely bound Cs atoms are surrounded by [Sn\textsubscript{2}Se\textsubscript{6}]/[Ge\textsubscript{2}Se\textsubscript{6}] units thereby exhibiting the same rattling-like effect, which can result in a common strong anharmonicity in both compounds.\cite{chang2018anharmoncity, li2022high}

{\color{blue}{Figures 5e and S4}} show the frequency dependence of the group velocity $\upsilon$ from all the \textit{q} points in different axes at 300 K for Cs\textsubscript{2}SnSe\textsubscript{3} and Cs\textsubscript{2}GeSe\textsubscript{3}, as given by $\upsilon_\lambda = {\partial\omega_\lambda}/{\partial q}$. Most of the phonon modes possess ultralow $\upsilon$ values less than 2 km/s, confirming the lattice softening induced by the weak interatomic bonding, which is also consistent with the flat phonon bands observed in the phonon dispersion. 
{\color{blue}{Figure 5f}} depicts the overall short $\tau$ values mainly ranging from 0.2 ps to 20 ps, with low-frequency phonons exhibiting a gradual shortening trend in both compounds. This can be attributed to the presence of LLO modes, which facilitate more scattering paths and impede the heat flow, significantly enhancing the scattering rates thereby reducing $\tau$ for both acoustic and LLO modes, ultimately reducing $\kappa_\mathrm{L}$. 
It is noteworthy that in the low-frequency phonon region, particularly within the acoustic phonon range, the phonon group velocities of these two compounds show little difference, yet the phonon lifetime of Cs\textsubscript{2}SnSe\textsubscript{3} is significantly longer than that of Cs\textsubscript{2}GeSe\textsubscript{3}. Therefore, we conclude that the relatively lower anomalous $\kappa_\mathrm{L}$ of Cs\textsubscript{2}GeSe\textsubscript{3} is attributed to its shorter phonon lifetime. 

We  speculate that the difference in the phonon lifetime between Cs\textsubscript{2}SnSe\textsubscript{3} and Cs\textsubscript{2}GeSe\textsubscript{3} might be related to the strength of Sn–Se and Ge–Se covalent bonding. This covalent bonding may induce anisotropic motion, involving the collective movement of Sn/Ge and Se atoms, thereby leading to a strong lattice anharmonicity within the system.\cite{kawano2021effect} 
To support this conclusion, we perform crystal orbital Hamilton population (COHP) calculations\cite{dronskowski1993crystal} to identify the energy-resolved local bonding information. {\color{blue}{Figure 5g}} shows that anti-bonding states persist down to –3 eV below the Fermi level ($E$\textsubscript{F}) in Sn–Se bonding, which can weaken the corresponding bonding strength thereby forming a polar covalent bond as observed in ELF. 
This can be demonstrated by the integral of COHP ($\mathrm{ICOHP}=\int_{-\infty}^E\mathrm{COHP}(E)\mathrm{d}E$) at the $E$\textsubscript{F}, which represents all occupied orbitals, serving as an indicator of bond strength. Since the average ICOHP values for Sn–Se and Ge–Se bonds are –0.72 eV and –4.49 eV, respectively, indicating the obviously weaker Sn–Se bonding in Cs\textsubscript{2}SnSe\textsubscript{3}. The stronger bonding strength of Ge–Se is also evident from its larger force constant $\left|\Phi_{ij}\right|$ in comparison to that of Sn–Se ({\color{blue}{Figure S5}}).
Hence, the shorter phonon lifetime of Cs\textsubscript{2}GeSe\textsubscript{3} relative to Cs\textsubscript{2}SnSe\textsubscript{3} may be attributed to their significant difference in Ge–Se and Sn–Se covalent bonding, which can be further quantified at the structural descriptor level through subsequent interpretable ML approaches. 

\subsection{Interpretable Supervised Classification for Predicting Ultralow $\kappa_\mathrm{L}$}
\begin{figure*}
\centering
\includegraphics[width=1\textwidth]{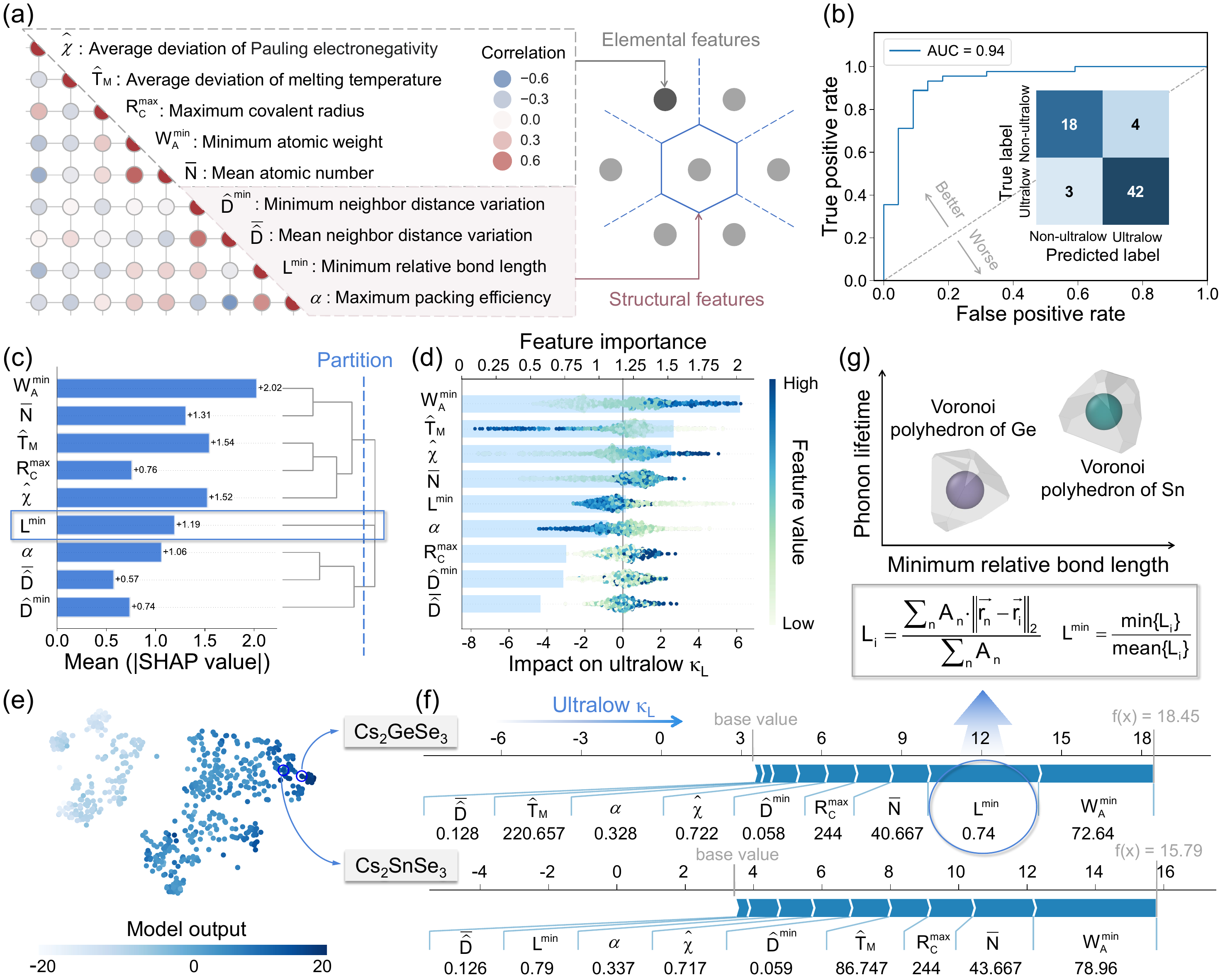}
\caption{(a–g) Supervised learning based on the local database and interpretable analysis. (a) Heat map of the Pearson correlation coefficient matrix among the selected descriptors and their definitions, which are classified into elemental and structural features. (b) ROC curve and confusion matrix of the optimal ML classification model for predicting ultralow/non-ultralow $\kappa_\mathrm{L}$. (c) Hierarchical clustering of the descriptors based on their SHAP values. (d) Global interpretability into $\kappa_\mathrm{L}$ classification using SHAP, including the feature importance and the impact of different features on ultralow $\kappa_\mathrm{L}$. (e) Using t-SNE visualization for the overall contributions of all descriptors of each material to the model output based on their SHAP values. (f) Local interpretability into $\kappa_\mathrm{L}$ classification using SHAP, including the force plots of Cs\textsubscript{2}GeSe\textsubscript{3} and Cs\textsubscript{2}SnSe\textsubscript{3}. (g) Discovered special descriptor $L^{\mathrm{min}}$ and its trend in influencing the phonon lifetime of Cs\textsubscript{2}GeSe\textsubscript{3} and Cs\textsubscript{2}SnSe\textsubscript{3}.} 

\end{figure*}

After labeling the materials in the second-level dataset based on our HTC framework to obtain the local database, we develop interpretable supervised classification models to predict ultralow $\kappa_\mathrm{L}$ for efficiently by-passing the complex \textit{ab initio }calculations, which can further refine the results of unsupervised learning and provide greater robustness. Here, materials with $\kappa_\mathrm{PET}$ not exceeding 2 W/mK are labeled as 1, which are considered to possess ultralow $\kappa_\mathrm{L}$; otherwise, they are labeled as 0, signifying non-ultralow $\kappa_\mathrm{L}$. 

Beyond the input dataset itself, identifying the most relevant features and appropriate algorithms can significantly enhance the generalization and accuracy of ML models, which are also quite strongly intertwined.
The chemical compositions and crystal structures of these compounds are also featurized into 273 descriptors based on the aforementioned Magpie data and Voronoi tessellations. As for the ML models, we compare eight widely-used classification algorithms for predicting material properties according to the indicators of Area Under the Receiver Operating Characteristic Curve (ROC AUC) and accuracy based on stratified ten-fold cross-validation ({\color{blue}{Figure S6}}): Decision Tree (DT), Extra Trees (ET), Random Forest (RF), Gradient Boosting Classifier (GBC), Adaptive Boosting (AdaBoost), eXtreme Gradient Boosting  (XGBoost), Light Gradient Boosting Machine (LightGBM), Categorical Boosting (CatBoost). Most algorithms demonstrate good performances, with the CatBoost algorithm achieving the best results, which is chosen to build the follow-up classification model. To avoid the curse of dimensionality, we perform feature selection using the model-based wrapper method and Pearson correlation method. Descriptors are iteratively removed based on the feature importance scores obtained from the CatBoost algorithm, and an ML model is subsequently trained using the remaining descriptors at each step. This process yields the curve depicting the model performance as a function of the number of features ({\color{blue}{Figure S7}}), indicating that optimal performance is achieved when utilizing the top 11 features. To reduce the correlation between features, the Pearson correlation coefficients (\textit{p}) of these 11 features are calculated, which are defined as: $p={\operatorname{cov}(x_i,x_j)}/{\sigma_{x_i}\sigma_{x_j}}$, where $\operatorname{cov}(x_i,x_j)$ is the covariance of features $x_i$ and $x_j$, $\sigma_{x_{i/j}}$ is the standard deviation of the feature $x_{i/j}$. For pairs of features with $\left| p \right|$ \textgreater\ 0.8, we retain the one with the higher feature importance score. {\color{blue}{Figure 6a}} displays the heatmap of the Pearson correlation coefficient matrix among the selected nine features and illustrates the specific meanings of these features based on elemental attributes and Wigner-Seitz cells, indicating that we have successfully eliminated redundant descriptors. We adjust the hyperparameters of the ML model based on Bayesian optimization over 200 random trials, obtaining the best hyperparameters as listed in {\color{blue}{Table S4}}. As a result, the performances of the optimal ML classification model is characterized by the ROC curve and the confusion matrix ({\color{blue}{Figure 6b}}), showing its great ability in classifying semiconductors with ultralow $\kappa_\mathrm{L}$ due to the high ROC AUC (0.94), accuracy (0.90), precision (0.90), recall (0.90) and F-score (0.90).

To establish the bridge between descriptors and thermal transport mechanisms thereby demystifying the black-box nature of ML, we perform interpretable analysis for the pre-trained $\kappa_\mathrm{L}$ classification model based on the SHAP method.\cite{shapley1953value, lundberg2017adv} 
{\color{blue}{Figure 6c}} shows the dendrogram of the optimal nine descriptors via using the hierarchical clustering based on their SHAP values, where a partition line classifies these features into three groups. Interestingly, beyond the minimum relative bond length $L^{\mathrm{min}}$, two classes of descriptors based on chemical compositions and crystal structures are identified as we expected. This indicates that $L^{\mathrm{min}}$ may possess a unique influence on ultralow $\kappa_\mathrm{L}$. 

The feature importance and the impact of different features on ultralow $\kappa_\mathrm{L}$ are shown in {\color{blue}{Figure 6d}}, which make a global interpretation for the classification model. We here take the most important feature in the three classes as an example to detailedly explain the corresponding hierarchical influence on $\kappa_\mathrm{L}$ combining the SHAP values. 
For the most important feature $ W_{\mathrm{A}}^{\mathrm{min}}$, defined as:
\begin{equation}
     W_{\mathrm{A}}^{\mathrm{min}} = \mathrm{min} \{  W_{\mathrm{A}}^{\mathrm{1}}, ..., W_{\mathrm{A}}^i, ..., W_{\mathrm{A}}^n \}
\end{equation}
Where $W_{\mathrm{A}}^i$ is the atomic weight of element \textit{i} and \textit{n} is the number of elements, respectively. Semiconductors with larger $ W_{\mathrm{A}}^{\mathrm{min}}$ values tend to exhibit low $\kappa_\mathrm{L}$ since the corresponding positive SHAP values, whereas those with smaller $ W_{\mathrm{A}}^{\mathrm{min}}$ values show the opposite trend. This can be explained by the dispersion relation for the frequency, just as $\omega=2\sqrt{{\beta}/M}\left|\sin{qa}/2\right|$ in a one-dimensional crystal lattice, where $\beta$ and $M$ are the bond force constant and the atomic mass, respectively. Large $M$ ($W$) can reduce the frequency $\omega$ and thus the group velocity $\upsilon$, resulting in a low $\kappa_\mathrm{L}$. 
For the feature $\alpha$, defined as:
\begin{equation}
    \alpha=\frac{NV_\mathrm{max}}{V}\times100
\end{equation}
Where $N$ is the number of atoms, $V_\mathrm{max}$ is the largest sphere volume occupied by one atom that can fit inside its Voronoi cell, $V$ is the cell volume. It is obvious that low $\alpha$ tends to have a positive effect on ultralow $\kappa_\mathrm{L}$ since its positive SHAP values, which can be explained by its direct effect on the bond length. $\alpha$ helps in understanding how closely the atoms are bonded, with lower values often signifying larger void space with longer bond lengths in a crystal structure. Longer bond lengths typically result in softer lattices characterized by lower group velocity $\upsilon$, thereby reducing $\kappa_\mathrm{L}$.
As for the special feature $L^{\mathrm{min}}$, defined as:
\begin{equation}
    L^{\min}=\frac{\mathrm{min}\{L_i\}}{\mathrm{mean}\{L_i\}}
\end{equation}
Where $L_i$ is the weighted average bond length of atom \textit{i} in a crystal structure, defined as:
\begin{equation}
    L_i=\frac{\sum_nA_n\cdot\left\|\overrightarrow{r_n}-\overrightarrow{r_i}\right\|_2}{\sum_nA_n}
\end{equation}
Where $\overrightarrow{r_i}$ is the position of atom \textit{i}, $\overrightarrow{r_n}$ and $A_n$ is the position and the area of the $n^{\mathrm{th}}$ neighbor of atom \textit{i}, respectively. $L_i$ provides a comprehensive description of the interactions between each atom and its nearest neighboring atoms in the crystal structure, with the result influenced by two components: the distance term $\left\|\overrightarrow{r_n}-\overrightarrow{r_i}\right\|_2$ and the weighted term $A_n$. The distance term $\left\|\overrightarrow{r_n}-\overrightarrow{r_i}\right\|_2$ directly governs the bond length between the nearest neighboring atoms, with shorter distances implying stronger bonding, thus leading to a lower $L_i$. The weighted term $A_n$ assigns varying weights to different nearest neighboring atoms, where the bond lengths corresponding to the nearest neighboring atoms with greater weights show a higher effect on $L_i$. 
Therefore, $\mathrm{min}\{L_i\}$ captures the local information within the structure, which may correspond to covalent bonds due to their typical shorter bond lengths. In contrast, $\mathrm{mean}\{L_i\}$ captures the global information, encompassing both short covalent bonds, long ionic bonds, etc. We propose that the ratio of these two terms, i.e., $L^{\mathrm{min}}$, might potentially reflect the anharmonicity of materials. 
Lower $L^{\mathrm{min}}$ values exhibit positive SHAP values, which favor the emergence of ultralow $\kappa_\mathrm{L}$. This implies that when a structure features lower $\mathrm{min}\{L_i\}$ and higher $\mathrm{mean}\{L_i\}$, it may possess rigidly distorted and weakly bound units with a significant bonding hierarchy, typically leading to stronger anharmonicity, as previously analyzed. 

The overall contributions of all descriptors of each material to the model
output based on their SHAP values are visualized by the t-SNE method ({\color{blue}{Figure 6e}}), showing that ultralow $\kappa_\mathrm{L}$ (dark blue) and non-ultralow $\kappa_\mathrm{L}$ (light blue) are clearly clustered into the corresponding groups. Cs\textsubscript{2}GeSe\textsubscript{3} and Cs\textsubscript{2}SnSe\textsubscript{3} are then marked out together to reveal the local interpretability using the SHAP force plots ({\color{blue}{Figure 6f}}), thereby exploring their previously discussed unusual difference of $\kappa_\mathrm{L}$. Among the features that have the most significant impact on these two materials,  $L^{\mathrm{min}}$ surprisingly reaches a level of importance in Cs\textsubscript{2}GeSe\textsubscript{3} comparable to that of $ W_{\mathrm{A}}^{\mathrm{min}}$. Although $ W_{\mathrm{A}}^{\mathrm{min}}$ in Cs\textsubscript{2}GeSe\textsubscript{3} is lower than in Cs\textsubscript{2}SnSe\textsubscript{3}, the crucial role of $L^{\mathrm{min}}$ in anharmonicity drives the predicted $\kappa_\mathrm{L}$ of Cs\textsubscript{2}GeSe\textsubscript{3} to be much lower. 
Given the significant difference in phonon lifetime $\tau$ between these two materials, as indicated by previous phonon thermal transport analysis, we speculate that $L^{\mathrm{min}}$ might have a great influence on $\tau$, which can be supported by two perspectives. 
On the one hand, the trend between $L^{\mathrm{min}}$ and $\tau$ in Cs\textsubscript{2}GeSe\textsubscript{3} and Cs\textsubscript{2}SnSe\textsubscript{3} indicates that the former not only exhibits a lower $L^{\mathrm{min}}$ but also has a significantly shorter $\tau$ ({\color{blue}{Figure 6g}}). This aligns with the impact of $L^{\mathrm{min}}$ on anharmonicity revealed by the SHAP analysis. 
On the other hand, {\color{blue}{Table S5}} indicates that the $\mathrm{min}\{L_i\}$ values in Cs\textsubscript{2}GeSe\textsubscript{3} and Cs\textsubscript{2}SnSe\textsubscript{3} correspond to the $L$ values derived from the Voronoi polyhedra centered on Ge and Sn, respectively. Although Cs\textsubscript{2}GeSe\textsubscript{3} has a lower $\mathrm{mean}\{L_i\}$ compared to Cs\textsubscript{2}SnSe\textsubscript{3}, its much lower $\mathrm{min}\{L_i\}$ (i.e., $L_{\mathrm{Ge}}$) eventually results in a much lower $L^{\mathrm{min}}$. This is consistent with the results based on first-principles analysis, which attributes the difference in phonon lifetime to the stronger bond strength (shorter bond length) of Ge–Se covalent bonding compared to that of Sn–Se.


\section{Conclusion}
In summary, we propose a novel HiBoFL framework via integrating unsupervised learning and supervised learning to efficiently predict complex properties and uncover structure-property relationships. As a compelling demonstration, this framework has been applied to efficiently identify semiconductors with ultralow $\kappa_\mathrm{L}$, which circumvents large-scale brute-force \textit{ab initio} calculations without clear objectives. By employing unsupervised learning for the materials from the MP database, we successfully group them into seven clusters. A few hundred materials in clusters \textit{C}\textsubscript{1} and \textit{C}\textsubscript{2} with a high likelihood of possessing low $\kappa_\mathrm{L}$ is selected from a pool of hundreds of thousands based on similarity design rules. We further conduct low-cost HTC on materials belonging to the two clusters, establishing a local database for researchers to retrieve and providing a list of candidate materials with potential applications in the TE field. Additionally, statistical analysis of these candidates offers valuable domain knowledge to guide the design of materials with ultralow $\kappa_\mathrm{L}$. Cs\textsubscript{2}GeSe\textsubscript{3} and Cs\textsubscript{2}SnSe\textsubscript{3} with ultralow $\kappa_\mathrm{L}$  ($ \sim $0.25 W/mK) are screened out, in which the anomalous ultralow $\kappa_\mathrm{L}$ of Cs\textsubscript{2}GeSe\textsubscript{3} is attributed to the lower $\tau$ caused by the difference in covalent bonding. Based on the established local database, we train a robust ML classification model to refine unsupervised learning, achieving a ROC AUC of 0.94 on the test set, which can enable the efficient predictions of ultralow $\kappa_\mathrm{L}$ materials. The interpretable analysis of the classification model reveals the mechanisms of key descriptors on $\kappa_\mathrm{L}$ modulation, such as $ W_{\mathrm{A}}^{\mathrm{min}}$, $\alpha$ and $L^{\mathrm{min}}$, etc. The special factor $L^{\mathrm{min}}$ is discovered to show a unique influence on structural anharmonicity, leading to the difference of phonon lifetime in Cs\textsubscript{2}GeSe\textsubscript{3} and Cs\textsubscript{2}SnSe\textsubscript{3}, in agreement with the first-principles analysis. We believe that this work provides a novel feasible way for efficiently seeking promising TE materials, in which the proposed HiBoFL framework is also expected to be applied in other material fields.

\section{Methods}
\subsection{First-Principles Calculations}
All the involved DFT-based first-principles calculations were carried out by using the projector-augmented wave (PAW) method\cite{kresse1999ultrasoft} to deal with ion-electron interactions as implemented in the Vienna \textit{Ab initio} Simulation Package (VASP),\cite{kresse1996efficient} in which the processing of data was conducted using VASPKIT.\cite{wang2021vaspkit} The electronic exchange-correlation energy was described by the Perdew–Burke–Ernzerhof (PBE) functional under the generalized gradient approximation (GGA).\cite{perdew1996generalized} Our automatic HTC workflow conducted on the candidate structures identified by unsupervised learning was accomplished within the framework of Python Materials Genomics (Pymatgen).\cite{ong2013python} With a plane-wave kinetic energy cutoff of 520 eV and a $\Gamma$-centered \textit{k}-point grid of 2$\pi$ $\times$ 0.04 $\mathrm{Å^{-1}}$ to sample the Brillouin zone, the structure optimization was terminated when the total energy convergence reached below $10^{-6}$ eV and the norms of all the forces were less than 0.01 eV/Å. The elastic properties of each material were calculated by applying a 1\% change to the volume of the optimized conventional cell. To evaluate the dynamic stability and extract the second-order interaction-force constants (2nd-order IFCs) of Cs\textsubscript{2}SnSe\textsubscript{3} and Cs\textsubscript{2}GeSe\textsubscript{3}, we used the finite displacement method\cite{baroni2001phonons} for calculations as implemented in the Phonopy package.\cite{phonopy-phono3py-JPCM, phonopy-phono3py-JPSJ} The obtained 2nd-order IFCs were utilized to construct the dynamic matrix and compute the corresponding harmonic properties. Additionally, we utilized the script thirdorder.py\cite{PhysRevB.86.174307} to generate the 2 $\times$ 2 $\times$ 1 and 2 $\times$ 2 $\times$ 1 supercells in consideration of the $10^\mathrm{th}$ nearest neighbors for Cs\textsubscript{2}SnSe\textsubscript{3} and Cs\textsubscript{2}GeSe\textsubscript{3}, thereby resulting in 1504 and 1500 supercells with displaced atoms for self-consistent calculations, respectively. The third-order interaction force constants (3rd-order IFCs) were further extracted to obtain the three-phonon scattering matrix elements, facilitating the calculations of anharmonic properties. Ultimately, we obtained the convergent $\kappa_\mathrm{L}$ values of these selected materials as  implemented in the ShengBTE package\cite{ShengBTE_2014} within a  20 $\times$ 20 $\times$ 20 \textit{q}-point grid in reciprocal space. The crystal structures and ELF diagrams were visualized using the Crystal Toolkit\cite{horton2023crystal} and VESTA.\cite{momma2011vesta} Moreover, we calculated the COHP as implemented in the LOBSTER code\cite{dronskowski1993crystal} to identify the bonding characteristics as bonding, anti-bonding, or non-bonding.

\subsection{Theoretical Framework }
Taking into account the extremely high cost of precisely calculating $\kappa_\mathrm{L}$, we employed the PET model proposed in our previous work\cite{yan2022high} within the HTC framework to obtain the corresponding values at 300 K for materials identified by unsupervised learning. The PET model has established the relationship between intrinsic $\kappa_\mathrm{L}$ and elastic properties, considering both acoustic phonon and optical phonon contributions, which achieves a certain balance between accuracy and efficiency. The empirical equation of $\kappa_\mathrm{PET}$ based on the PET model is expressed as:
    \begin{equation}\
    \begin{split}
        \kappa_\mathrm{PET}=\frac{(6\pi^{2})^{2/3}}{3\pi^{3}}\cdot\frac{\overline{M}}{T(\overline{V}N)^{2/3}}\frac{\overline\upsilon^{3}}{\overline{\gamma}^{2}}+\frac{3k_{\mathrm{B}}\overline\upsilon}{2\overline{V}^{2/3}}\Big(\frac{\pi}{6}\Big)^{1/3}(1-N^{-2/3})
    \end{split}
    \end{equation}
Where $\overline{M}$ is the average atomic mass, \textit{T} is the temperature, $\overline{V}$ is the average atomic volume, \textit{N} is the number of atoms in the primitive cell, $k_\mathrm{B}$ is the Boltzmann constant, $\overline\upsilon$ and $\overline{\gamma}$ are the average sound velocity and average Grüneisen parameter, respectively. Among them, we can obtain $\overline\upsilon$ as given by:
\begin{equation}
    \begin{aligned}\overline\upsilon& = [\frac{1}{3}(\overline\upsilon_{\mathrm{l}}^{-1} + 2\overline\upsilon_{\mathrm{t}}^{-3})]^{-1/3}\\& 
    = \{\frac{1}{3}(\frac{\overline{M}}{\overline{V}})^{3/2}[(B + \frac{4}{3}G)^{-3/2} + 2G^{-3/2}]\}^{-1/3}\end{aligned}
\end{equation}
And $\overline{\gamma}$ can be expressed as:
\begin{equation}
    \begin{aligned}
      \overline\gamma&=\sqrt{\frac13(\overline\gamma_\mathrm{l}^2+2\overline\gamma_\mathrm{t}^2)} \\&
      =\sqrt{\frac13 \{[\frac12\frac{\partial\mathrm{ln}\left(B+\frac43G\right)}{\partial\mathrm{ln}V}+\frac16]^2 + 2(\frac12\frac{\partial\mathrm{ln~}G}{\partial\mathrm{ln~}V}+\frac16)^2\}}
    \end{aligned} 
\end{equation}
Where $\overline\upsilon_{\mathrm{l}}$, $\overline\upsilon_{\mathrm{t}}$, $\overline\gamma_\mathrm{l}$ and $\overline\gamma_\mathrm{t}$ are the longitudinal sound velocity, transverse sound velocity, longitudinal Grüneisen parameter, and transverse Grüneisen parameter, respectively. \textit{B} and \textit{G} are bulk moduli and shear moduli, respectively. In this manner, we can calculate the $\kappa_\mathrm{PET}$ for different materials derived from their elastic properties within the HTC framework. 

To further validate $\kappa_\mathrm{PET}$ from the PET empirical equation and investigate the phonon thermal transport mechanisms, we calculated more accurate $\kappa_\mathrm{L}$ values for Cs\textsubscript{2}SnSe\textsubscript{3} and Cs\textsubscript{2}SnSe\textsubscript{3} by iteratively solving the phonon BTE:\cite{ShengBTE_2014, fugallo2013ab}
\begin{equation}
    \kappa_\mathrm{L}^{\alpha\beta}=\frac{1}{k_\mathrm{B}T^2\Omega N}\sum_\lambda f_0\left(f_0+1\right)\left(\hbar\omega_\lambda\right)^2\upsilon_\lambda^\alpha F_\lambda^\beta
\end{equation}
Where $\alpha$ and $\beta$ are the Cartesian indexes. $k_\mathrm{B}$, $T$, $\Omega$, $N$ are the Boltzmann constant, temperature, volume of the unit cell, and regular grid of \textit{q} points, respectively. $\lambda$ is a phonon mode including the branch index \textit{p} and wave vector \textit{q}, and $f_0$ is the phonon distribution function based on Bose-Einstein statistics. $\hbar$, $\omega_\lambda$, $\upsilon_\lambda^\alpha$ are the reduced Planck constant, phonon frequency, and phonon group velocity along the $\alpha$ direction, respectively. When only considering two- and three-phonon processes that contribute to scattering, the linearized BTE $F_\lambda$ takes the form: 
\begin{equation}
    F_\lambda=\tau_\lambda^0(\upsilon_\lambda+\Delta_\lambda)
\end{equation}
Where $\tau_\lambda^0$ and $\Delta_\lambda$ are the phonon lifetime of mode $\lambda$ and corrective term from iteration, respectively.

\subsection{Machine Learning}
The chemical compositions and crystal structures of materials were featurized into different descriptors using the Matminer package.\cite{ward2018matminer} All parts related to ML were carried out using the Scikit-learn package.\cite{pedregosa2011scikit} To achieve the automation and acceleration of hyperparameter optimization in supervised learning, we employed the powerful Optuna package\cite{akiba2019optuna} for efficiently finding the best hyperparameters. And to provide the interpretable analysis of the black-box ML model, a game-theoretic approach as implemented in the SHAP package\cite{shapley1953value, lundberg2017adv, lundberg2020local} was employed.

For unsupervised learning, the elbow method and the silhouette coefficient method are utilized to determine the optimal \textit{k} value in \textit{k}-means clustering. In the elbow method, the sum of squared errors (SSE) is defined as:
\begin{equation}
    SSE=\sum_{i=1}^k\sum_{x\in C_i}\|x-\mu_i\|^2
\end{equation}
Where $C_i$ represents the set of data points in cluster \textit{i}, $x$ is a data point in $C_i$, $\mu_i$ is the centroid of $C_i$. The silhouette coefficient is defined as:
\begin{equation}
    s(x)=\frac{b(x)-a(x)}{\max(a(x),b(x))}
\end{equation}
Where $b(x)$ is the average intra-cluster distance and measures how close the data point \textit{x} is to other points in $C_i$, $a(x)$ is the average nearest-cluster distance and measures how far data point \textit{x} is from the closest other cluster. After calculating the silhouette coefficient for all data points, the average value is then taken to obtain the mean silhouette coefficient for different values of \textit{k}. 

For supervised learning, accuracy, precision, recall, F-score, and ROC AUC are utilized to evaluate the performance metrics of classification models, which are defined as:
\begin{equation}
    Accuracy=\frac{TP+TN}{TP+TN+FP+FN}
\end{equation}
\begin{equation}
    Precision=\frac{TP}{TP+FP}
\end{equation}
\begin{equation}
    Recall=\frac{TP}{TP+FN}
\end{equation}
\begin{equation}
    F\text{-}socre=\frac{2TP}{2TP+FP+FN}
\end{equation}
\begin{equation}
   ROC\ AUC = \int_{0}^{1} TPR \, d(FPR)
\end{equation}
Where TP is the number of true positives, FP is the number of false positives, TN is the number of true negatives, and FN is the number of false negatives. Moreover, TPR and FPR are the true positive rate and false positive rate, respectively, which are defined as: 
\begin{equation}
   TPR=\frac{TP}{TP+FN}
\end{equation}
\begin{equation}
   FPR=\frac{FP}{TN+FP}
\end{equation}

For interpretable analysis, the SHAP value of a material descriptor is given by:
\begin{equation}
    \varphi_i(p) = \sum_{S \subseteq F \setminus \{i\}} \frac{|S|! (|F| - |S| - 1)!}{|F|!} \left( p(S \cup \{i\}) - p(S) \right)
\end{equation}
Where $p$ is the prediction model, $F$ represents the set of all material descriptors, $S$ represents the subset of all descriptors excluding the feature \textit{i}, $ p(S)$ is the model prediction when only descriptors in $S$ are considered, and $ p(S \cup \{i\})$ is the model prediction when feature \textit{i} is also included.

\acknowledgements
We acknowledge the support from the National Natural Science Foundation of China (No. 11935010), the National Key R\&D Program of China (No. 2023YFA1406900 and No. 2022YFA1404400), the Natural Science Foundation of Shanghai (No. 23ZR1481200), the Program of Shanghai Academic Research Leader (No. 23XD1423800), and the Opening Project of Shanghai Key Laboratory of Special Artificial Microstructure Materials and Technology.

\bibliography{mybib}

\end{document}